\documentclass[aps,prd,reprint,superscriptaddress,nofootinbib,longbibliography]{revtex4-1}
\usepackage{amsmath,amssymb, amsthm,amstext}
\usepackage{natbib}
\usepackage{graphicx}
\usepackage{color}
\usepackage{array, enumerate}
\usepackage{bm}
\usepackage{multirow}
\usepackage[breaklinks,colorlinks,citecolor=blue]{hyperref}
\usepackage{braket}
\usepackage{txfonts}
\usepackage{physics}
\usepackage[normalem]{ulem}
\usepackage{float}
\usepackage{subcaption}
\usepackage[dvipsnames]{xcolor}
\usepackage[singlelinecheck=false]{caption}

\makeatletter
\renewcommand{\p@subtable}{}
\makeatother

\def\nn{\nonumber}
\def\be{\begin{equation}}
\def\ee{\end{equation}}
\newcommand{\comment}[1]{}

\begin{document}

\title{Quasinormal Mode Excitation in the Horizon Shear of Near-Extremal Kerr Black Holes}

\author{Brennen Black}
\email{brennenblack@sjtu.edu.cn}
\affiliation{Tsung-Dao Lee Institute, Shanghai Jiao-Tong University, Shanghai, 520 Shengrong Road, 201210, China}
\affiliation{School of Physics \& Astronomy, Shanghai Jiao-Tong University, Shanghai, 800 Dongchuan Road, 200240, China}

\begin{abstract}
The horizon shear of a near-extremal Kerr black hole perturbed by an eccentric extreme mass ratio inspiral shows low amplitude, high frequency oscillations (''wiggles") after periapsis passage. Previous works are unable to tie the frequency of the wiggles to the orbit, and argue they cannot be quasinormal modes (QNMs). We show that they are Kerr QNMs belonging to the damped-mode branch --- that is, modes whose decay rate stays finite as the spin approaches extremality. For an equatorial prograde orbit with $p = 3M$ and $e = 0.85$ and spins ranging from $a/M = 0.999$ to $a/M = 0.99999$, fits to the time-domain shear identify the fundamental $(\ell,m) = (2,1)$ and $(3,2)$ modes as the dominant and first subdominant excitations. Fitting each multipole separately gives frequencies to within two standard deviations of the Kerr values and provides evidence for the excitation of the fundamental $(3,1)$, $(4,1)$ and $(4,2)$ modes, though these are not resolvable in the full signal. We further argue that the same wiggles must be present in the tidal field that drives the shear. For $\ell \leq 4$ we find no evidence of QNM excitation in the $m = 0$ or the $m = \ell$ multipoles.
\end{abstract}
\date{\today}

\maketitle

\section{\label{sec:level1}Introduction}
Extreme Mass Ratio Inspirals (EMRIs) are a deeply studied source of gravitational waves. These systems are, in their most general state, comprised of a Kerr black hole primary of mass $M$ and angular momentum $J$ along with a point mass orbiter. In the literature, the primary is often described using the angular momentum per unit mass $a = J/Mc$ or the dimensionless spin $ac^{2}/GM$. With the upcoming space-based gravitational wave detector, the Laser Interferometer Space Antenna (LISA), much of the gravitational wave literature has been focused on measurements and phenomena at infinity \cite{Barack2004, Babak2017, AmaroSeoane2007}. Even from the early stages, a significant portion of the literature is focused on efficient construction of accurate EMRI waveforms \cite{Chua2021,Katz2021,Hughes2021,Isoyama2022,Wardell2023,Burke2024}. After waveform generation was established, some works began exploring how environmental factors around the primary (for example, thin accretion disks or gas disks) affect the measured gravitational waves, even going as far as to predict environmental parameters from the observed waveform \cite{Kocsis2011,Barausse2014,Speri2023}. While not as predominant as gravitational radiation at infinity, a number of works have considered how an EMRI system influences the primary black hole, which we refer to as the horizon side. In 1972, Stephen Hawking and James Hartle derived the laws governing how a Kerr hole's mass and spin change under external radiation \cite{Hawking1972}, laying the foundation for studying horizon absorption, tidal effects and late time tails in an EMRI setting \cite{Hartle1973,Hartle1974,Teukolsky1974,Poisson1995,Tagoshi1997,Hughes2001,Alvi2001,Poisson2004}.

Of interest to this paper are the many studies on tidal deformation of the black hole horizon. Such work originated in the context of tides in neutron stars (for example, see \cite{Venumadhav2013,Binnington2009}), but was quickly expanded to black hole horizons \cite{Vega2011,Taylor2008,Damour2009,Poisson2010}. \cite{OSullivan2014,OSullivan2016,Penna2017} calculate and visualize the horizon deformation for Kerr EMRIs, \textcolor{black}{and \cite{OSullivan2016} in particular discovered an unusual phenomenon. When examining the near-extremal spin regime, they noticed damped, high-frequency, low-amplitude wiggles persisting after the initial burst signal. The authors were unable to relate the wiggles' frequency to the orbit, and previous computational difficulties has made it difficult to attribute this phenomenon to Kerr QNM excitations. The authors note that while these wiggles are present in the horizon shear, they are not present in the tidal driving response and thereby conclude they cannot be Kerr QNMs.}

In 2020, \cite{Thornburg2020} \textcolor{black}{conducted an in-depth study of this wiggle phenomenon.} They showed that these wiggles were actually a common feature in the EMRI gravitational waves (and related quantities) at infinity \textcolor{black}{and were able to relate them to Kerr QNMs at low and moderate spins.} They showed that the wiggles were present for a significant portion of the parameter space; that is, the wiggles were not necessarily a near-extremal phenomenon. Importantly they showed that these wiggles were not resonances by showing that the excitation amplitude remained finite as the driving frequency swept past the QNM frequency; \textcolor{black}{this phenomenon was therefore classified as QNM excitation as opposed to a resonant phenomenon.}

These results leave an open question. \cite{OSullivan2016} argue the horizon wiggles are not QNMs because none were identified in the tidal field, whereas \cite{Thornburg2020} showed that the analogous wiggles at infinity are QNM excitations. In this paper we return to investigate the wiggles in the horizon shear to determine whether they are Kerr QNMs. Answering this requires accurate Kerr QNMs at near-extremal spin, which recent methods now provide \cite{Cook2014, Stein2019, Casals2019, CarneirodaCunha2021} (see Section~\ref{sec:level2.4}). Kerr QNMs, the source-free solutions of the Teukolsky equation \cite{Teukolsky1973, Press1973}, are central to black hole spectroscopy \cite{Berti2006a} and have been computed analytically and numerically across most of the spin range \cite{Detweiler1980, Leaver1985, Seidel1990, Onozawa1997, Berti2006a, Dolan2010, Cook2014, Cook2016}. As the spin approaches extremality, the spectrum of multipoles with $m/(\ell+\tfrac{1}{2}) \lesssim 0.74$ splits into two branches \cite{Yang2012, Yang2013a, Yang2013b, Cook2014, Casals2019, CarneirodaCunha2021}: zero-damping modes, whose decay rate tends to zero, and damped modes, whose decay rate stays finite. We test whether the horizon-shear wiggles are QNM excitations by fitting a sum of QNMs and a polynomial background to the time-domain shear and measuring its agreement with Kerr QNMs using the Mahalanobis distance \cite{Mahalanobis2018}.

The paper is structured as follows. In Section \ref{sec:level2} we provide a brief theoretical review of the Kerr spacetime, calculating the ingoing gravitational radiation and the horizon deformation, and provide a more detailed discussion on computing Kerr QNMs in the near-extremal spin regime. In Section \ref{sec:level3.1} we generate the full time-domain horizon deformation data. We perform a standard least squares fit using a QNM-plus-polynomial-background model in order to extract the two dominant modes, and compare with the Kerr QNM spectrum. Motivated by the results of Section \ref{sec:level3.1}, in Section \ref{sec:level3.2} we look to isolate each mode to allow for more accurate identification of each mode. In Section \ref{sec:level3.3} we then compare our results from Sections \ref{sec:level3.1} and \ref{sec:level3.2}, then provide interpretation and further discussion. We then conclude in Section \ref{sec:level4}.

\section{\label{sec:level2}Theory Review}
In this section, we provide a brief review of the basic theory of horizon-side gravitational waves and the horizon shear response. A thorough review is contained in \cite{OSullivan2014,OSullivan2016,Penna2017}, so here we present enough theory for the paper to remain self-contained. Throughout this paper we use the conventions that $G = c = 1$. \textcolor{black}{In written equations we leave the Kerr black hole mass $M$ as a free parameter, but in computations take $M = 1$.}\\
\subsection{\label{sec:level2.1}The Spacetime}
The Kerr solution describes the spacetime of a massive rotating black hole with spin parameter $a$. The Kerr metric is typically presented in Boyer-Lindquist (BL) coordinates $(t,r,\theta,\phi)$ with Lorentzian signature
\begin{align}\label{eq:KerrMetricBL}
    ds^2 = &-\left(1 - \frac{2Mr}{\Sigma}\right)dt^2 - \frac{4Mar\sin^{2}(\theta)}{\Sigma}dtd\phi + \frac{\Sigma}{\Delta}dr^2\nn\\
    &+ \Sigma d\theta^2 + \frac{(r^2+a^2)^2-a^2\Delta\sin^{2}(\theta)}{\Sigma}\sin^{2}(\theta)d\phi^2,
\end{align}
where
\begin{equation}\label{eq:SigmaDefn}
    \Sigma = r^2 + a^2\cos^{2}(\theta)
\end{equation}
and
\begin{equation}\label{eq:DeltaDefn}
    \Delta = r^2 - 2Mr + a^2.
\end{equation}
$\Delta$ has two roots, $r_{\pm} = M \pm \sqrt{M^2 - a^2}$ which correspond to the inner (-) and outer (+) horizons.

The outer horizon is of particular interest, and we will benefit from a representation of the spacetime that is well behaved at $r = r_{+}$. To achieve this requires two transformations. The first is a coordinate transformation, moving from BL coordinates to Kerr ingoing (KI) coordinates using
\begin{subequations}\label{eq:BLtoKIDiffTrans}
    \begin{align}
        dv &= dt + \frac{r^2+a^2}{\Delta}dr,\label{eq:BLtoKIdv}\\
        d\psi &= d\phi + \frac{a}{\Delta}dr\label{eq:BLtoKIdpsi}.
    \end{align}
\end{subequations}
The above coordinate transformation removes the coordinate singularity in Eq.\eqref{eq:KerrMetricBL}, yielding a metric that is not singular at the outer horizon $r_{+}$. One can also integrate the transformations in Eq.\eqref{eq:BLtoKIDiffTrans} to find
\begin{subequations}\label{eq:BLtoKITrans}
    \begin{align}
        v &= t + \underbrace{r + \frac{Mr_{+}}{\sqrt{M^2-a^2}}\ln\left(\frac{r}{r_{+}}-1\right) - \frac{Mr_{-}}{\sqrt{M^2-a^2}}\ln\left(\frac{r}{r_{-}}-1\right)}_{\equiv r^{*}},\label{eq:BLtoKIv}\\
        \psi &= \phi + \underbrace{\frac{a}{2\sqrt{M^2-a^2}}\ln\left(\frac{r-r_{+}}{r-r_{-}}\right)}_{\equiv \bar{r}}.\label{eq:BLtoKIpsi}
    \end{align}
\end{subequations}
In particular, one recognizes the r-dependent terms of Eq.\eqref{eq:BLtoKIv} are the tortoise coordinate in the Kerr spacetime, making $v = t + r^{*}$ the advanced time. The advantage of using these coordinates may not immediately be clear, as Eq.\eqref{eq:BLtoKITrans} clearly diverges at $r = r_{+}$. However, as explained in \cite{OSullivan2014,OSullivan2016,Penna2017} the combination $\bar{r} - \Omega_{H}r^{*}$, where $\Omega_{H} = a/2Mr_{+}$ is the horizon's angular frequency, converges to a finite number in the limit that $r\to r_{+}$
\begin{align}\label{eq:KaDefn}
    K(a) \equiv &\lim_{r\to r^{+}}\bar{r}-\Omega_{H}r^{*} = \frac{a}{2M(Mr_{+}-a^2)}\Bigg\{a^2 - Mr_{+} \nonumber\\
    &+ M^2\ln\left(\frac{r_{+}}{r_{-}}\right) + M\sqrt{M^2-a^2}\ln\left(\frac{a^2}{4(M^2-a^2)}\right)\Bigg\}.
\end{align}
\textcolor{black}{In this way, the divergences brought on by the $\ln(r-r_{+})$ terms in Eqs.\eqref{eq:BLtoKIv} and \eqref{eq:BLtoKIpsi} to the combination $\bar{r} - \Omega_{H}r^{*}$ cancel each other in the horizon limit, leaving a function $K(a)$ that depends only on the spin parameter $a$. This result is important for the discussion of Section \ref{sec:level2.3}, in particular regarding the phase of the gravitational wave solution, which contains the term $\bar{r} - \Omega_{H}r^{*}$. Its finite value on the horizon is crucial to ensuring the derived solutions are indeed well-behaved on the horizon.}
\subsection{\label{sec:level2.2}Ingoing Gravitational Radiation}
\indent The ingoing gravitational radiation is contained within the complex scalar $\psi_0$ of the Newman-Penrose formalism. This scalar is a first-order gauge invariant quantity of spin weight $s = +2$, and its evolution is governed by the Teukolsky equations \cite{Teukolsky1973}. It is given by the projection of the Weyl tensor $C_{\mu\nu\sigma\lambda}$ onto the $\ell$ and $m$ tetrad legs
\begin{equation}\label{eq:Psi0Defn}
    \psi_0 = -C_{\mu\nu\sigma\lambda}\ell^{\mu}m^{\nu}\ell^{\sigma}m^{\lambda}.
\end{equation}
The Teukolsky equation is typically presented in the Kinnersley tetrad, but this choice makes $\psi_0$ singular at the horizon. As one approaches the horizon $\psi_0$ diverges as $\Delta^{-2}$ in the Kinnersley tetrad. The second transformation required is therefore a tetrad transformation, more specifically a transformation to the Hawking-Hartle (HH) tetrad \cite{Hawking1972,Teukolsky1973,Chandrasekhar:1983}. For a more detailed account of the Kinnersley tetrad, the HH tetrad, and the transformation between them we refer readers to the appendix of \cite{OSullivan2014}. One determines a very simple relation between $\psi_0$ in the Kinnersley tetrad ($\psi_{0}^{\rm K}$) and in the HH tetrad ($\psi_{0}^{\rm HH}$) \textcolor{black}{[cf. Ref. \cite{OSullivan2014}, Eq. (2.9)]}
\begin{equation}\label{eq:KtoHHforPsi0}
    \psi_{0}^{\rm HH}(r) = \left[\frac{\Delta}{2(r^2+a^2)}\right]^2\psi_{0}^{\rm K}(r).
\end{equation}

Importantly, one sees that the near-horizon $\Delta^{-2}$ behavior \textcolor{black}{of $\psi_{0}^{\rm K}$ cancels with the $\Delta^2$ in Eq.\eqref{eq:KtoHHforPsi0}}, indicating regularity of $\psi_{0}^{\rm HH}$ on the horizon.\\
\indent To obtain a solution for $\psi_{0}^{\rm K}$ (and therefore by Eq.\eqref{eq:KtoHHforPsi0} a solution for $\psi_{0}^{\rm HH}$) one performs a Fourier transform in time $t$, then applies a separation of variables in the radial and angular variables. For an EMRI on a stable bound orbit and starting in BL coordinates, one writes
\begin{equation}\label{eq:BL-psi0K}
    \psi_{0}^{\rm K}(t,r,\theta,\phi) = \sum_{\ell mkn}{}_{+2}R^{\rm inhomo}_{\ell mkn}(r){}_{+2}S_{\ell mkn}(\theta;a\omega_{mkn})e^{i(m\phi - \omega_{mkn}t)},
\end{equation}
where
\begin{equation}\label{eq:GWfreqs}
    \omega_{mkn} = m\Omega_{\phi} + k\Omega_{\theta} + n\Omega_{r}, \quad m\in\mathbb{Z}_{[-\ell, \ell]},\quad k,n\in\mathbb{Z}.
\end{equation}
The term ${}_{+2}R^{\rm inhomo}_{\ell mkn}(r)$ is the inhomogeneous solution to the radial Teukolsky equation \textcolor{black}{[cf. Eq.(3) of Ref. \cite{Lo:2025lpo}]}
\begin{align}\label{eq:InhomoRadialSolution}
    {}_{+2}R^{\rm inhomo}_{\ell m\omega}(r) &= \frac{{}_{+2}R^{\rm up}_{\ell m\omega}(r)}{W_{R}}\int_{r_{+}}^{r}\frac{{}_{+2}R^{\rm in}_{\ell m\omega}(\tilde{r}){}_{+2}\mathcal{T}_{\ell m\omega}(\tilde{r})}{\Delta^{-2}(\tilde{r})}d\tilde{r} \nonumber\\
    &+ \frac{{}_{+2}R^{\rm in}_{\ell m\omega}(r)}{W_{R}}\int_{r}^{\infty}\frac{{}_{+2}R^{\rm up}_{\ell m\omega}(\tilde{r}){}_{+2}\mathcal{T}_{\ell m\omega}(\tilde{r})}{\Delta^{-2}(\tilde{r})}d\tilde{r},
\end{align}
in which $W_{R}$ is the Wronskian of the two homogeneous solutions ${}_{+2}R^{\rm in,up}_{\ell m\omega}$ (which satisfy the usual QNM boundary conditions at the horizon and infinity, respectively), and ${}_{+2}\mathcal{T}_{\ell m\omega}$ is the EMRI source term that drives the Teukolsky equation. ${}_{+2}S_{\ell mkn}(\theta;a\omega_{mkn})$ is the solution to the angular Teukolsky equation, referred to as spin-weighted spheroidal harmonics. Both of these can be obtained numerically using the open-source Generalized Sasaki-Nakamura (GSN) code released by Rico Lo and collaborators \cite{Lo:2023fvv,Yin:2025kls,Lo:2025lpo}. The GSN code contains a built-in Teukolsky formalism solver, which is non-divergent when calculating EMRI waveforms because the Teukolsky source term is not extended. The frequencies $\Omega_{\phi},\Omega_{\theta},\Omega_{r}$ characterize the orbit of the EMRI, and are dependent on the primary's spin parameter $a$, the orbit's semi-latus rectum $p$ and eccentricity $e$, along with the cosine of the orbital inclination (denoted in this work by $x = \cos(\theta_{\rm inc})$) \cite{Schmidt2002,Mino2003,Fujita2009}. These fundamental frequencies can be calculated using the \textsc{KerrGeoPy} python package \cite{kerrgeopy}. The fundamental frequencies play a particularly important role because, as is implied by Eq.\eqref{eq:GWfreqs}, the Fourier transform only has support at a discrete lattice of frequencies \cite{Drasco2004,Drasco2006}. Combined with the spheroidal decomposition, the solution becomes a discrete sum over the four summation indices $\ell,m,k,n$, where $\ell \geq 2$, $-\ell \leq m \leq \ell$ and $k,n\in\mathbb{Z}$.

When we take $r\to r_{+}$, ${}_{+2}R^{\rm inhomo}_{\ell mkn}(r)$ \textcolor{black}{is proportional to the ingoing homogeneous solution ${}_{+2}R^{\rm in}_{\ell mkn}(r)$} \cite{Lo:2023fvv,Yin:2025kls,Lo:2025lpo}. \textcolor{black}{We denote the (complex-valued) constant of proportionality by ${}_{+2}R^{\rm H}_{\ell mkn}$, and note that since ${}_{+2}R^{\rm in}_{\ell mkn}(r)\to\Delta^{-2}e^{-ip_{mkn}r^{*}}$ as $r\to r_{+}$ \cite{Lo:2023fvv,Yin:2025kls} we have that}
\begin{equation}\label{eq:RinhomoAsymptote}
    {}_{+2}R^{\rm inhomo}_{\ell mkn}(r\to r_{+}) = {}_{+2}R^{\rm H}_{\ell mkn}\Delta^{-2}e^{-ip_{mkn}r^{*}},
\end{equation}
where $p_{mkn} = \omega_{mkn} - m\Omega_{H}$ is the effective frequency \textcolor{black}{of a given $(m,k,n)$ mode} in the black hole's co-rotating frame. The term ${}_{+2}R^{\rm H}_{\ell mkn}$ is the complex-valued horizon amplitude, which can be obtained using the GSN code with an EMRI as the source. Mathematically the horizon amplitude is obtained by taking the limit as $r\to r_{+}$ of Eq.~\eqref{eq:InhomoRadialSolution}. While the first integral in Eq~\eqref{eq:InhomoRadialSolution} vanishes, the second does not; that integral from $r_{+}$ to infinity and with the division by the Wronskian is what defines ${}_{+2}R^{\rm H}_{\ell mkn}$.

With all this information, we now make the transformation from BL coordinates to KI coordinates, and take the limit $r\to r_{+}$. Making use of Eq.\eqref{eq:KaDefn} and Eq.\eqref{eq:RinhomoAsymptote}, we find
\begin{equation}\label{eq:KI-HorizonPsi0K}
    \psi_{0}^{K}(v, r\to r_{+},\theta,\psi) = \Delta^{-2}\sum_{\ell mkn}{}_{+2}R^{\rm H}_{\ell mkn}{}_{+2}S_{\ell mkn}(\theta;a\omega_{mkn})e^{i\Phi(v,\psi)}
\end{equation}
with
\begin{align}\label{eq:KIPhase}
    \Phi(v,\psi) &= \lim_{r\to r_{+}}(m\phi - \omega_{mkn}t - p_{mkn}r^{*})\nonumber\\
    &= \lim_{r\to r_{+}}(m(\psi - \bar{r}) - \omega_{mkn}(v-r^{*}) - (\omega_{mkn} - m\Omega_{H})r^{*})\nonumber\\
    &= m\left(\psi - \lim_{r\to r_{+}}(\bar{r} - \Omega_{H}r^{*})\right) - \omega_{mkn}v \nonumber\\
    &= m(\psi - K(a)) - \omega_{mkn}v.
\end{align}
Finally, we obtain a solution for $\psi_{0}^{\rm HH}$ using Eq.\eqref{eq:KtoHHforPsi0} and taking $r \to r_{+}$
\begin{align}\label{eq:Psi0HHSolution}
    \psi_{0}^{\rm HH}(v,\theta,\psi) &= \frac{1}{4(r_{+}^{2} + a^2)^2}\sum_{\ell mkn}{}_{+2}R^{\rm H}_{\ell mkn}{}_{+2}S_{\ell mkn}(\theta;a\omega_{mkn})e^{i\Phi(v,\psi)}\nn\\
    &\equiv \sum_{\ell mkn}\psi_{0,\ell mkn}{}_{+2}S_{\ell mkn}(\theta;a\omega_{mkn})e^{i\Phi(v,\psi)}.
\end{align}
The ingoing gravitational radiation on the horizon can further be decomposed into two polarizations, the plus polarization $\psi_{0,+}^{\rm HH}$ and the cross polarization $\psi_{0,\times}^{\rm HH}$. Following \cite{OSullivan2014,OSullivan2016,Penna2017}, these are related to $\psi_0^{\rm HH}$ in a simple way
\begin{equation}\label{eq:Psi0HHPolarizations}
    \psi_0^{\rm HH}(v,\theta,\psi) = -(\psi_{0,+}^{\rm HH}(v,\theta,\psi) + i\psi_{0,\times}^{\rm HH}(v,\theta,\psi)).
\end{equation}
A special case arises when $(\theta,\psi) = (\pi/2, 0)$, in which the cross polarization identically vanishes on the horizon.
\subsection{\label{sec:level2.3}Horizon Shear}
Of the 12 spin coefficients in the Newman-Penrose formalism, the one of interest in horizon deformation is the shear, denoted by $\sigma$, and given at zeroth-order by
\begin{equation}\label{eq:ShearDefn}
    \sigma = \pm m^{\mu}m^{\nu}\nabla_{\mu}\ell_{\nu}.
\end{equation}
where the sign depends on the conventions, $\ell^{\mu}$ and $m^{\mu}$ are two of the four tetrad legs in the Newman-Penrose formalism, and $\nabla_{\mu}$ denotes standard covariant differentiation. At zeroth-order the shear vanishes as a consequence of the Kerr spacetime being Petrov type D \cite{Chandrasekhar:1983}, but at first-order it is in general non-vanishing. Typically, a quantity that vanishes at zeroth-order does not carry a first-order identifier. In other words, we follow the convention that $\sigma$ is already a first-order quantity, and abstain from differentiating its zeroth-order part from its first-order part since there is no room for confusion. Its equation of motion is obtained by straightforward manipulations of the Newman-Penrose field equations. On the horizon, one determines the equation to be \textcolor{black}{[cf. Ref. \cite{OSullivan2016} Eq.(2.32) and surrounding discussion]}
\begin{equation}\label{eq:ShearEOM}
    (\partial_{v} + \Omega_{H}\partial_{\psi} - \kappa(a))\sigma = \psi_{0}^{\rm HH}
\end{equation}
in the HH tetrad, where $\kappa(a) = \sqrt{M^2-a^2}/2Mr_{+}$ is the surface gravity (not to be confused with the spin-coefficient $\kappa$, which is nowhere relevant in this work). The role of $\psi_{0}^{\rm HH}$ is clear by the structure of Eq.\eqref{eq:ShearEOM}: it acts as the driver to the horizon deformation. To obtain a solution for $\sigma$, we look for solutions following the same decomposition as $\psi_0^{\rm HH}$
\begin{equation}\label{eq:ShearSolution}
    \sigma(v,\theta,\psi) = \sum_{\ell mkn}\sigma_{\ell mkn}{}_{+2}S_{\ell mkn}(\theta;a\omega_{mkn})e^{i\Phi(v,\psi)}.
\end{equation}
Plugging Eq.\eqref{eq:ShearSolution} into Eq.\eqref{eq:ShearEOM}, one finds that the shear ansatz satisfies the equation of motion so long as
\begin{equation}\label{eq:ShearTransferFunction}
    \sigma_{\ell mkn} = \frac{i}{p_{mkn} - i\kappa(a)}\psi_{0,\ell mkn}.
\end{equation}
Throughout this paper, we refer to the prefactor of Eq.\eqref{eq:ShearTransferFunction} as the shear transfer function. The simple algebraic relation between the $\sigma$ and $\psi_0^{\rm HH}$ amplitudes elucidates the teleological nature of the black hole horizon through the induced phase shift \cite{OSullivan2014,OSullivan2016,Penna2017}. Classically one expects that a driven oscillator cannot respond to an impulse before it arrives, but the horizon shear response can lead the driving gravitational field. As the boundary of the causal past of the future conformal infinity, the structure of the hole's horizon requires knowledge of the spacetime at future conformal infinity, and thus can evolve in a seemingly preemptive fashion \cite{Gourgoulhon_BH}. Furthermore, Eq.\eqref{eq:ShearTransferFunction} makes clear that the horizon shear admits the same polarization decomposition as $\psi^{\rm HH}_{0}$, namely that $\sigma(v,\theta,\psi) = \sigma_{+}(v,\theta,\psi) + i\sigma_{\times}(v,\theta,\psi)$. This sign difference between the polarization decompositions for $\sigma$ and $\psi_{0}^{\rm HH}$ are inherited from \cite{OSullivan2014,OSullivan2016,Penna2017} in order to remain consistent with their work.

\subsection{\label{sec:level2.4}Near-Extremal Kerr Quasinormal Modes}
Determining Kerr QNMs in the near-extremal limit requires much higher precision \textcolor{black}{than the typical distance between Kerr QNMs at low to moderate spins. It has been shown that in the near-extremal spin regime the Kerr QNM frequencies begin to bunch up against the real axis in the complex plane \cite{Yang2013a,Yang2013b}. As these poles inch closer to each other, root finding algorithms can spontaneously switch between two poles in close proximity.} In recent years, there have been many published works exploring new methods for maintaining accuracy and precision in the near-extremal regime (for some examples, see \cite{Casals2019,CarneirodaCunha2021}). In this work, we rely on the method of \cite{Cook2014} which has been implemented in the \textsc{qnm} python package \cite{Stein2019}. Table \ref{tab:RefKerrQNMs} shows the reference Kerr QNM frequencies obtained by the $\textsc{qnm}$ package truncated to four decimal points. In the cases where \textcolor{black}{QNM solutions exist but cannot be reliably identified}, we write a dash (---). \textcolor{black}{We discuss in more detail these shortcomings at the end of this section.} Furthermore, as will be discussed in Section \ref{sec:level3.3}, we only analyze up to $\ell = 4$ in this work.

\begin{table*}[t]
    \centering
    \begin{subtable}{\textwidth}
        \centering
        \renewcommand{\arraystretch}{1.3}
        \setlength{\tabcolsep}{10pt}
        \begin{tabular}{c|c|c|c|c|c}
            \hline
            \hline
            $a/M$ & $\omega_{210}$ & $\omega_{310}$ & $\omega_{320}$ & $\omega_{410}$ & $\omega_{420}$ \\
            \hline
            0.999 & $0.5806 - 0.0392i$ & $0.7947-0.0593i$ & $1.0256 - 0.0219i$ & $1.0127-0.0669i$ & $1.1909-0.0455i$ \\
            \hline
            0.9995 & $0.5810 - 0.0387i$ & $0.7950-0.0591i$ & $1.0272 - 0.0203i$ & $1.0130-0.0668i$ & $1.1917-0.0451i$ \\
            \hline
            0.9999 & $0.5814 - 0.0383i$ & $0.7952-0.0590i$ & $1.0283 - 0.0189i$ & $1.0132-0.0667i$ & $1.1923-0.0448i$ \\
            \hline
            0.99995 & $0.5814 - 0.0383i$ & $0.7952-0.0590i$ & $1.0284 - 0.0187i$ & $1.0132-0.0667i$ & $1.1924-0.0447i$ \\
            \hline
            0.99999 & $0.5814 - 0.0383i$ & $0.7952-0.0590i$ & $1.0285 - 0.0186i$ & $1.0132-0.0667i$ & $1.1925-0.0447i$ \\
            \hline
            \hline
        \end{tabular}
        \caption{A table of the fundamental ($n_{o} = 0$) Kerr QNM DM frequencies $\omega_{\ell mn_{o}}$ for the $(\ell,m) = (2,1), (3,1), (3,2), (4,1)$ and $(4,2)$ modes as a function of $a/M$, truncated to four decimals.}
        \label{tab:RKQ-A}
    \end{subtable}
    
    \vspace{1em}
    
    \begin{subtable}{\textwidth}
        \centering
        \renewcommand{\arraystretch}{1.3}
        \setlength{\tabcolsep}{10pt}
        \begin{tabular}{c|c|c|c|c|c}
            \hline
            \hline
            $a/M$ & $\omega_{211}$ & $\omega_{311}$ & $\omega_{321}$ & $\omega_{411}$ & $\omega_{421}$ \\
            \hline
            0.999 & $0.5397-0.1196i$ & $0.7735-0.1797i$ & --- & $0.9979-0.2018i$ & $1.1687-0.1372i$ \\
            \hline
            0.9995 & $0.5393-0.1191i$ & $0.7736-0.1794i$ & --- & $0.9981-0.2015i$ & $1.1690-0.1363i$ \\
            \hline
            0.9999 & $0.5389-0.1187i$ & $0.7737-0.1791i$ & --- & $0.9983-0.2013i$ & $1.1692-0.1356i$ \\
            \hline
            0.99995 & $0.5389-0.1187i$ & $0.7737-0.1790i$ & --- & $0.9983-0.2013i$ & $1.1692-0.1355i$ \\
            \hline
            0.99999 & $0.5389-0.1186i$ & $0.7737-0.1790i$ & --- & $0.9983-0.2013i$ & $1.1692-0.1355i$ \\
            \hline
            \hline
        \end{tabular}
        \caption{A table of the first overtone ($n_{o} = 1$) Kerr QNM DM frequencies $\omega_{\ell mn_{o}}$ for the $(\ell,m) = (2,1), (3,1), (4,1)$ and $(4,2)$ modes as a function of $a/M$, truncated to four decimals. We are unable to obtain reliable frequencies for the $(\ell, m, n_{o}) = (3,2,1)$ branch \textcolor{black}{as the imaginary part sporadically jumps between different ZDM-like trajectories with increasing spin}.}
        \label{tab:RKQ-A}
    \end{subtable}
\caption{A table containing the reference Kerr QNMs obtained with the \textsc{qnm} python package \cite{Stein2019} as a function of the spin parameter $a$. For those DM frequencies that cannot be reliably obtained, we write a dash (---).}
\label{tab:RefKerrQNMs}
\end{table*}

In the near-extremal Kerr limit, it has been shown that the QNMs split into two branches, a branch where the decay time approaches infinity as $a/M$ approaches unity, and another branch where the damping time asymptotes to a finite number in the same $a/M$ limit. Modes belonging to the former branch are called zero-damping modes (ZDMs), while those belonging to the latter branch are called damped modes (DMs) \cite{Yang2012,Yang2013a,Yang2013b}. According to \cite{Hod2008,Hod2013,Yang2013a}, the ZDM frequencies have a simple analytic formula
\begin{equation}\label{eq:ZDMfreq}
    \omega_{\rm ZDM} \to m\Omega_{H} - i\left(n_{o} + \frac{1}{2} + |\delta|\right)\kappa(a)
\end{equation}
where $n_{o}$ is the overtone number \textcolor{black}{(starting at $n_{o} = 0$ for the fundamental mode, and denoted this way so as to avoid confusion with the harmonic index $n$ of Eq.~\eqref{eq:GWfreqs})}, $\kappa(a)$ is the surface gravity, $\Omega_{H}$ is the hole's angular frequency and $\delta$ is given by \cite{Yang2013a,Hod2008,Hod2013}
\begin{equation}\label{eq:DeltaSqr}
    \delta^2 = \frac{7m^2}{4} - \left(s + \frac{1}{2}\right)^2 - {}_{s}\mathcal{A}_{\ell m}\left(\omega = \frac{m}{2},\frac{a}{M}=1\right).
\end{equation}
The term ${}_{s}\mathcal{A}_{\ell m}(\omega=m/2,a=1)$ is related to the eigenvalue of the angular Teukolsky equation $\lambda$ by \cite{Yang2013b,Lo:2023fvv}
\begin{equation}\label{eq:AngTeulkolskyEigenvalue}
    \lambda = {}_{s}\mathcal{A}_{\ell m}\left(\omega = \frac{m}{2}, \frac{a}{M}=1\right) + \left(\frac{m}{2}\right)^2 - m^2.
\end{equation}
We note that in general $\delta$ is written to be included in the real part of the frequency, but when $\delta^2 < 0$ it contributes to the imaginary part as $i|\delta|$. For our multipoles of interest $\delta^2$ is always negative, so we opt to write $\delta$ directly into the imaginary part in Eq.\eqref{eq:ZDMfreq}.

The bifurcation of the Kerr QNM spectrum in the near-extremal limit provides some obstacles to our analysis. The \textsc{qnm} package cannot discriminate between the ZDM and DM branches. \textcolor{black}{The \textsc{QNM} algorithm builds each QNM starting from $a = 0$, where there is no spectrum bifurcation, up to $a_{\rm max} = 0.9995M$. Beyond the max spin, the seeds for the root-finder are extrapolated from the lower spin results. Whether the algorithm tracks a ZDM branch or a DM branch in the near-extremal spin regime therefore depends entirely on the solutions near the bifurcation point. With no way to control the choice of branch there will inevitably be cases in} which we are unable to recover \textcolor{black}{DM frequencies, which will be of particular interest to this work. To ensure that the results of the \textsc{qnm} package can be trusted, we plot each $(\ell,m,n_{o})$ mode as a function of spin and check that the imaginary parts tend to a finite, non-zero limit as the spin approaches unity.} From this we determine that we cannot obtain DM frequencies for $(2,2,0)$, $(3,3,0)$, $(4,3,0)$, $(4,4,0)$, $(2,2,1)$, $(3,2,1)$, $(3,3,1)$ and $(4,4,1)$. In the case of $m = \ell$, there simply does not exist a DM branch \cite{Yang2013a,Yang2013b}. In the remaining cases the imaginary parts are sporadic, frequently switching between \textcolor{black}{different trajectories which all appear to converge to zero as the dimensionless spin $a/M$ approaches unity}.

\section{\label{sec:level3}Numerical Results}
In this work we analyze five different near-extremal Kerr spacetimes, $a/M \in \{0.999, 0.9995, 0.9999, 0.99995, 0.99999\}$. For each spacetime we use the same orbital parameters, namely $p = 3.0M$, $e = 0.85$ and $x = 1.0$, which provide a good compromise between excitation amplitude and grid resolution in the frequency-domain. To be clear the time-domain data varies smoothly with the parameters $p,e$ so we've chosen the parameters which give the clearest data to fit. Furthermore, in each system we take the viewing angles to be $\theta = \pi/2$ and $\psi = 0$. This specific configuration has the benefit of an identically vanishing cross polarization; we therefore limit the analysis to the plus polarization. Throughout the analysis we show the $a/M = 0.999$ data unless otherwise stated, \textcolor{black}{since the data of interest don't vary significantly in this near-extremal spin regime. We also note that from here on we take the Kerr mass to be $M = 1$.}

To avoid confusing the horizon shear $\sigma$ with the standard error budget (also represented by $\sigma$), we opt to write $\varepsilon$ to denote the uncertainty in a measurement. The measurements obtained in this section have a covariance matrix $\Sigma$, which in this work includes both statistical and systematic uncertainties. For clarity, we take $\Sigma$ to be the sum of the parameter covariance of the fiducial fit and the systematic sample covariance. The components of the systematic sample covariance are discussed in Section \ref{sec:level3.1}. To quantify the $\varepsilon$-deviation of a set of parameters $\vec{x}$ of length $q$ \textcolor{black}{and covariance matrix $\Sigma$} from a known set of parameters $\vec{\mu}$, we start by calculating the square of the Mahalanobis distance defined by \cite{Mahalanobis2018}
\begin{equation}\label{eq:MahalDist}
    D^2_{q} = (\vec{x} - \vec{\mu})^{\rm T}\Sigma^{-1}(\vec{x} - \vec{\mu}),
\end{equation}
where the superscript T denotes transposition, and $\Sigma^{-1}$ is the inverse of $\Sigma$. Assuming $\vec{x} \sim \mathcal{N}(\mu, \Sigma)$ it can be shown that $D^2_{q}$ is just the sum of $q$ independent and identically distributed squared standard normals \cite{Melendez2019}; this is exactly the definition of the chi-squared distribution with $q$ degrees of freedom, denoted by $\chi^{2}_{q}$. Since the data are noise-free and $\Sigma$ is built largely from systematic variations, this is an approximation, and $n\varepsilon$ (as defined next) should be read as a nominal measure of agreement.

For a single Gaussian parameter, a deviation of $n$ standard deviations has a familiar meaning: a larger deviation occurs with probability $\alpha_n = \mathrm{erfc}(n/\sqrt{2})$. With several correlated parameters there is no single standard deviation, so we carry this meaning over through the probability. We say that $\vec{x}$ lies within $n\varepsilon$ of $\vec{\mu}$ if $D^2_q \leq D^2_{n\varepsilon}$, where the threshold $D^2_{n\varepsilon}$ is chosen so that $\chi^2_q$ exceeds $D^{2}_{n\varepsilon}$ with the same probability $\alpha_n$. For a fitted frequency $\vec{x} = (M\omega_R, M\omega_I)$ (so $q = 2$), $\chi^2_2$ exceeds $D^2_{n\varepsilon}$ with probability $\exp(-D^2_{n\varepsilon}/2)$, so $D^2_{n\varepsilon} = -2\ln\alpha_n = \{2.296, 6.180, 11.829\}$ for $n = \{1, 2, 3\}$ respectively. 

\subsection{\label{sec:level3.1}Time-Domain Fitting}
Using the GSN code \cite{Lo:2023fvv,Yin:2025kls,Lo:2025lpo}, we can efficiently calculate the complex-valued coefficient ${}_{+2}R^{\rm H}_{\ell mkn}{}_{+2}S_{\ell mkn}(\theta=\pi/2;a\omega_{mkn})$. Eq.\eqref{eq:ShearSolution} with amplitudes from Eq.\eqref{eq:Psi0HHSolution} and Eq.\eqref{eq:ShearTransferFunction} allow us to calculate the time-domain tidal driving field and horizon shear response. For each configuration, we calculate up to $\ell = 9$ and $-150 \leq n \leq 150$. For an equatorial orbit, the amplitude of the polar motion \textcolor{black}{(motion which changes the value of the polar angle $\theta$)} vanishes so a summation over $k$ \textcolor{black}{(that which was defined in Eq.\eqref{eq:GWfreqs})} is unnecessary --- in practice we set $k = 0$ for all calculations, and further omit $k$ from any labels.
\begin{figure}
    \centering
    \includegraphics[width=1.0\linewidth]{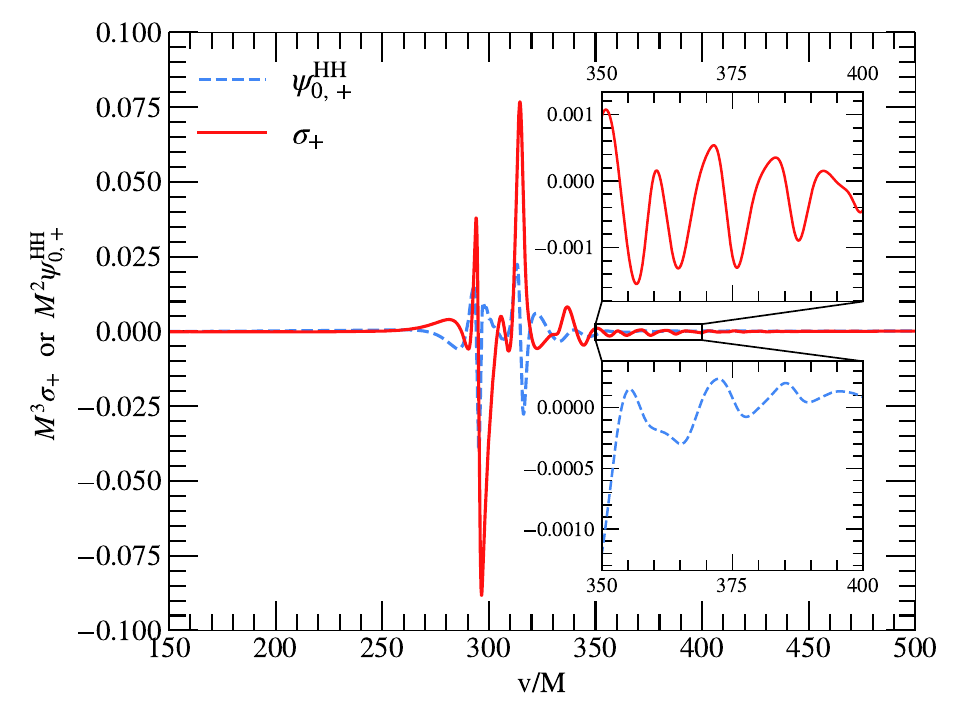}
    \caption{A plot of the tidal driving field $\psi_{0,+}^{\rm HH}$ (blue dashed line) and the horizon shear response $\sigma_{+}$ (solid red line) in the time domain, over roughly one orbit of semi-latus rectum $p = 3.0$, eccentricity $e = 0.85$, and inclination $x = 1.0$. The primary is taken to have $a/M = 0.999$, and the signal is viewed from an observer located at the viewing angles $(\theta, \psi) = (\pi/2, 0)$. With such a large eccentricity, the secondary spends a significant amount of time orbiting slowly near apoapsis ($v/M \sim 150-250$), and does not produce strong gravitational waves. As it falls towards and subsequently whips around periapsis ($v/M \sim 300$), the orbiter generates a burst-like gravitational wave signal. The low-amplitude high-frequency wiggles can be seen after roughly $v/M \sim 350$ where the prompt burst has significantly decayed. The top inset shows the start of the wiggle phenomenon in the horizon shear response, while the bottom inset shows the wiggles' existence in the tidal driving field. The difference in the wiggle structure and amplitude between the two insets is a result of the transfer function Eq.\eqref{eq:ShearTransferFunction}.}
    \label{fig:FiducialTDSignal}
\end{figure}
\begin{figure}
    \centering
    \includegraphics[width=1.0\linewidth]{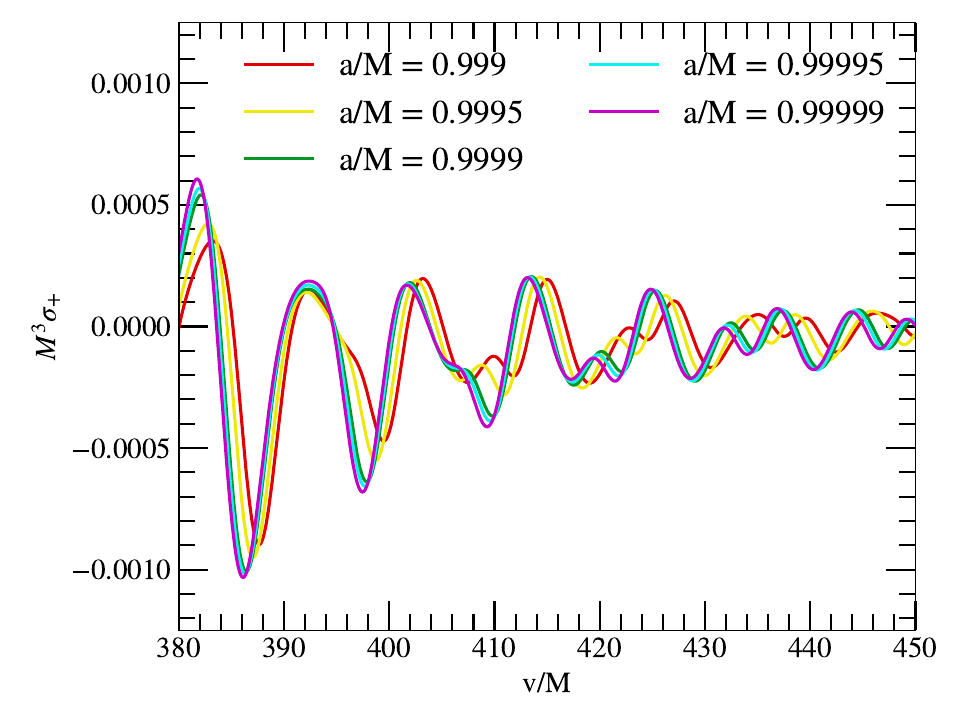}
    \caption{A zoomed in view of the horizon shear wiggles for each $a/M\in\{0.999, 0.9995, 0.9999, 0.99995, 0.99999\}$. All orbits use $p = 3.0$, $e = 0.85$ and $x = 1.0$ and are viewed from an observer located at $(\theta,\psi) = (\pi/2, 0)$.}
    \label{fig:TD_QNMperSpin}
\end{figure}

In Figure \ref{fig:FiducialTDSignal}, we show an example of the tidal driving field and the horizon shear response as seen in the time domain for $a/M = 0.999$. We do not consider circularization or inspiral effects due to gravitational radiation so each orbit produces roughly the same periodic signals (modulo precession effects which are included in the orbit). We restrict all plots and analysis to the first full-orbit burst. The dominant excitation in Figure \ref{fig:FiducialTDSignal} is nearly identical in structure and amplitude for all spins. In all configurations, the wiggles become visible after roughly $v/M = 350$ as the driven response fades.

In Figure \ref{fig:TD_QNMperSpin} we show a zoomed in view of the wiggles for all spins. A number of particularly important observations arise. First, the fact that the observed wiggles decay at all directs us to look for Kerr QNMs as opposed to the \textcolor{black}{zeros of the denominator} of Eq.\eqref{eq:ShearTransferFunction} \textcolor{black}{(which are also by definition poles, but lead to solutions growing in $v$ and so cannot be QNMs)}. Second, the wiggle structure is broadly consistent among all spins indicating that, if Kerr QNMs are indeed excited, the same $(\ell,m,n_{o})$ modes dominate the late time signal in each configuration. Last, the apparent decay time is roughly the same for all spins. This leads us to suspect that excited modes likely belong to the DM branch.

To extract the QNM frequencies in the wiggles, we perform a least squares fit with equal weights on all parameters. Our fitting model $\mathcal{M}$ sums an arbitrary number of complex exponentials with an arbitrary-degree polynomial background,
\begin{equation}\label{eq:FitModel}
    \mathcal{M} = \Re\left(\sum_{j}A_{j}e^{-i\omega_{j}(v-v_{\rm start})}\right) + \sum_{j}c_{j}(v-v_{\rm start})^{j},\quad v_{\rm start} \leq v \leq v_{\rm end}
\end{equation}
where $A_{j},\omega_{j}\in\mathbb{C}$ and $c_{j}\in\mathbb{R}$. In practice the most reliable results come from fitting two complex exponentials and a second-degree polynomial background. Since we don't know the exact mathematical form of the background, the polynomial background serves as an approximation, and is only valid in the decaying region of our data (where the wiggles are present). To ensure that the background doesn't fit to the wiggles we keep the degree as low as reasonably possible; we deem this to be quadratic order, hence the use of a second-degree polynomial background for all fiducial fits.

For all spins, we identify any $v_{\rm start}\in[365, 390]M$ and any $v_{\rm end}\in[475, 500]M$ as reasonable endpoints for the fitting window. To obtain our results, we simply choose the middle of each of these ranges; that is, the fiducial fitting window is chosen to be $v\in[377.5, 487.5]M$ and this yields the fitting results and statistical uncertainty. We perform a window sweep over the previously stated interval endpoints to generate the systematic uncertainty associated with the choice of the fitting interval. Furthermore, we vary the background polynomial degree to include the systematic uncertainty of our choice of background model. All uncertainties reported include both the statistical and systematic uncertainties.
\begin{table*}[t]
    \centering
    \begin{subtable}{\textwidth}
        \centering
        \renewcommand{\arraystretch}{1.3}
        \setlength{\tabcolsep}{10pt}
        \begin{tabular}{c|c|c|c|c|c|c|c|c}
            \hline \hline
            Spin & \multicolumn{4}{c|}{Mode 1} & \multicolumn{4}{c}{Mode 2} \\
            \hline
            $a/M$ & $M\omega_{R}$ & $M\varepsilon_{R}$ & $M\omega_{I}$ & $M\varepsilon_{I}$ & $M\omega_{R}$ & $M\varepsilon_{R}$ & $M\omega_{I}$ & $M\varepsilon_{I}$ \\
            \hline
            0.999 & 0.5804 & 0.0099 & -0.0467 & 0.0032 & 1.0249 & 0.0032 & -0.0229 & 0.0029 \\
            \hline
            0.9995 & 0.5794 & 0.0095 & -0.0461 & 0.0030 & 1.0280 & 0.0029 & -0.0187 & 0.0022 \\
            \hline
            0.9999 & 0.5777 & 0.0087 & -0.0450 & 0.0028 & 1.0322 & 0.0027 & -0.0130 & 0.0019 \\
            \hline
            0.99995 & 0.5772 & 0.0085 & -0.0448 & 0.0027 & 1.0332 & 0.0027 & -0.0112 & 0.0018 \\
            \hline
            0.99999 & 0.5766 & 0.0082 & -0.0447 & 0.0025 & 1.0352 & 0.0026 & -0.0106 & 0.0018 \\
            \hline
            \hline
        \end{tabular}
        \caption{A table detailing the fit results $(M\omega_{R},M\omega_{I})$ and their uncertainties $(M\varepsilon_{R},M\varepsilon_{I})$ for each fitted mode.}
        \label{tab:TDFit-A}
    \end{subtable}

    \vspace{1pt}

    \begin{subtable}{\textwidth}
        \centering
        \renewcommand{\arraystretch}{1.3}
        \setlength{\tabcolsep}{10pt}
        \begin{tabular}{c|c|c|c|c}
            \hline \hline
            Spin & \multicolumn{2}{c|}{Mode 1 $D_{2}^{2}$} & \multicolumn{2}{c}{Mode 2 $D_{2}^{2}$} \\
            \hline
            $a/M$ & $(\ell,m,n_{o})=(2,1,0)$ & $(\ell,m,n_{o})=(2,1,1)$ & $(\ell,m,n_{o})=(3,2,0)$ & $(\ell,m,n_{o})=(4,1,0)$ \\
            \hline
            0.999 & 5.4171 & 523 & 0.1256 & 244 \\
            \hline
            0.9995 & 5.9106 & 585 & 0.5956 & 485 \\
            \hline
            0.9999 & 5.8414 & 708 & 15.7419 & 1064 \\
            \hline
            0.99995 & 5.9409 & 753 & 28.1387 & 1280 \\
            \hline
            0.99999 & 6.5707 & 849 & 38.6045 & 1389 \\
            \hline
            \hline
        \end{tabular}
        \caption{We present two square Mahalanobis distances $D^{2}_{2}$ using the data provided in Table \ref{tab:TDFit-A} and the covariance matrices determined from the fitting process. In the first column of each mode, we provide the square Mahalanobis distance to the closest Kerr QNM from Table \ref{tab:RefKerrQNMs}. In the second column, we show the square Mahalanobis distance to the second closest Kerr QNM from Table \ref{tab:RefKerrQNMs}. The large difference in distance between the closest and second closest modes (the first and second columns of each mode at each $a/M$, respectively) makes clear that the mode assignment is unambiguous: each fitted mode is definitively closest to the first column's Kerr QNM, and there is no plausible alternative assignment.}
        \label{tab:TDFit-B}
    \end{subtable}
    \caption{A table containing the fit results and the Mahalanobis distances for each mode.}
    \label{tab:TDFitValues}
\end{table*}
\begin{figure}
    \centering
    \includegraphics[width=1.0\linewidth]{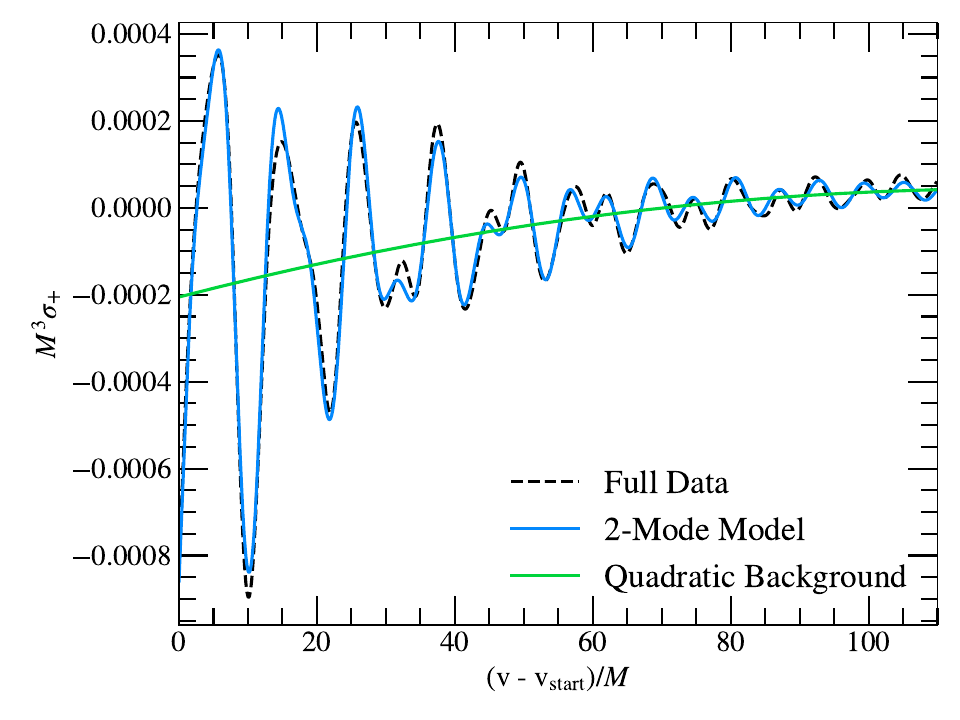}
    \caption{Here we show a fiducial example of the time-domain fitting results for $a/M = 0.999$ with the orbit characterized by $p = 3.0$, $e = 0.85$ and $x = 1.0$ as viewed from an observer located at $(\theta,\psi) = (\pi/2, 0)$. We plot the original data (black dashed line), the signal reconstructed from the fitted parameters (solid blue line), and the quadratic background (solid green line).}
    \label{fig:ExampleTDFit}
\end{figure}

In Figure \ref{fig:ExampleTDFit} we show an example of the time-domain fitting results for $a/M = 0.999$. The oscillatory structure is generally consistent between the full time-domain data and the signal reconstructed from the fitted parameters, but there is a \textcolor{black}{clear mismatch in the two signals}. \textcolor{black}{Given that our model in Eq.\eqref{eq:FitModel} is truncated to only two QNMs and a degree-2 polynomial this is not unexpected, but we should take care to confirm that the mismatch can} be mostly attributed to the background \textcolor{black}{as opposed to missing QNM content}. To test this, we considered fitting a third mode to the data. It was determined that the frequency varied widely with the start time --- the real part of the frequency varied between roughly 0.39 and 0.98, while the imaginary part of the frequency varied between -0.02 and -0.15 --- while the results for the first two modes remained roughly the same, indicating that this was not a mode detection. In each case, the added mode reduced the mismatch at early times but left the late-time deficit essentially unchanged. As will be discussed in a later section, the estimation of the amplitudes of the next-strongest modes lies below the two-mode residual, so any additional QNM content is unresolvable in our setup without first reducing the residuals. We therefore \textcolor{black}{are confident in attributing} the mismatch primarily to an imperfect background.

Table \ref{tab:TDFitValues} shows the fitted frequencies and their uncertainties for each mode and for each spin, along with the Mahalanobis distance $D_{2}^{2}$ to the nearest Kerr QNM and the next-nearest neighbor. For all spins, the first mode is closest to the $(2,1,0)$ Kerr QNM while the second mode is closest to the $(3,2,0)$ Kerr QNM. However, while the first mode stays within $3\varepsilon$ of the $(2,1,0)$ Kerr QNM, the second mode seems to \textcolor{black}{drift} away from the Kerr $(3,2,0)$ mode at $a/M \geq 0.9999$. When looking specifically at Table \ref{tab:TDFit-A}, we notice the real part of the second mode shifts by 0.0103 while the uncertainty drops from 0.0032 to 0.0026 as spin increases, which is a significant trend. A similar story is told for the imaginary part of the second mode, which shifts by 0.0123 while the uncertainty drops from 0.0029 at $a/M = 0.999$ to 0.0018 at $a/M = 0.99999$. Again, this trend is clearly resolvable within the error budget. For the moment, we defer discussion of this trend to Section \ref{sec:level3.3}.

\subsection{\label{sec:level3.2}Mode Identification and Isolation}
In the previous section we fit to the time-domain data, which includes the sum of multipoles up to and including $\ell = 9$. While we were able to obtain somewhat reasonable fit parameters, our identification of the excited Kerr QNMs could benefit from verification. One is therefore led to ask where in Eq.\eqref{eq:ShearSolution} QNM excitation lives, and it quickly becomes clear that the QNM excitation must live in the complex amplitude
\begin{equation}\label{eq:SigEllMNDefn}
    \sigma_{\ell mn} = \sigma_{\ell m0n}{}_{+2}S_{\ell m0n}\left(\theta = \frac{\pi}{2}; a\omega_{m0n}\right).
\end{equation}
\begin{figure}
    \centering
    \includegraphics[width=1.0\linewidth]{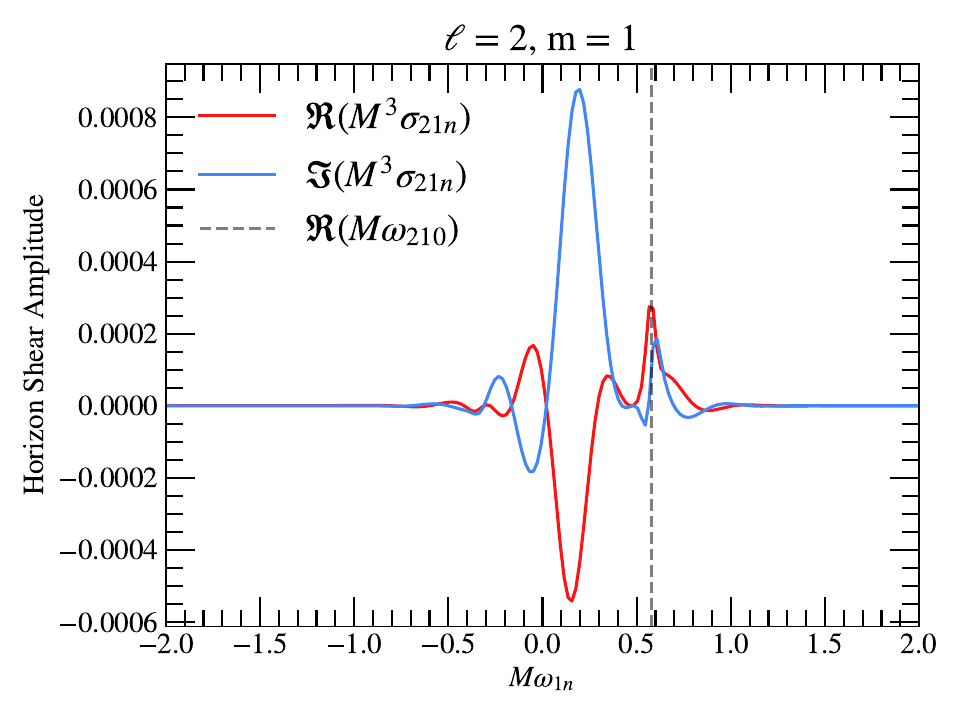}
    \caption{A plot of $M^{3}\sigma_{21n}$ as a function of $M\omega_{1n}$ for $a/M = 0.999$, $p = 3.0$, $e = 0.85$, $x = 1.0$ and $(\theta,\psi) = (\pi/2, 0)$. The gray dashed line shows the real part of the fundamental mode for $(\ell,m) = (2,1)$ at $a/M = 0.999$.}
    \label{fig:l2m1amp}
\end{figure}
\begin{figure}
    \centering
    \includegraphics[width=1.0\linewidth]{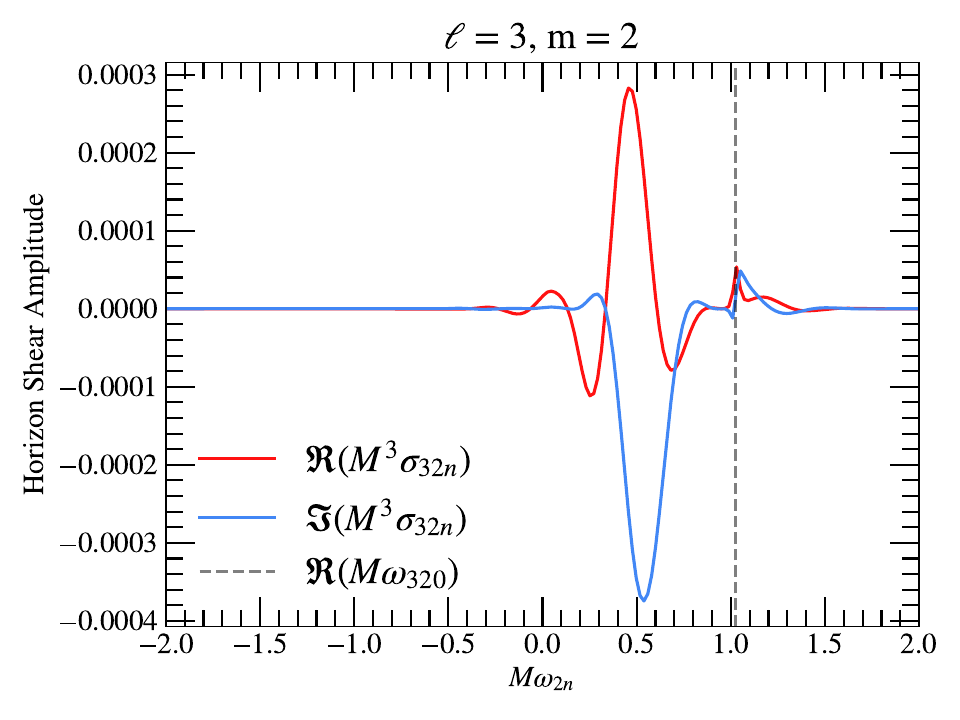}
    \caption{A plot of $M^{3}\sigma_{32n}$ as a function of $M\omega_{2n}$ for $a/M = 0.999$, $p = 3.0$, $e = 0.85$, $x = 1.0$ and $(\theta,\psi) = (\pi/2, 0)$. The gray dashed line shows the real part of the fundamental mode for $(\ell,m) = (3,2)$ at $a/M = 0.999$.}
    \label{fig:l3m2amp}
\end{figure}
\begin{figure*}
    \centering
    \includegraphics[width=1.0\linewidth]{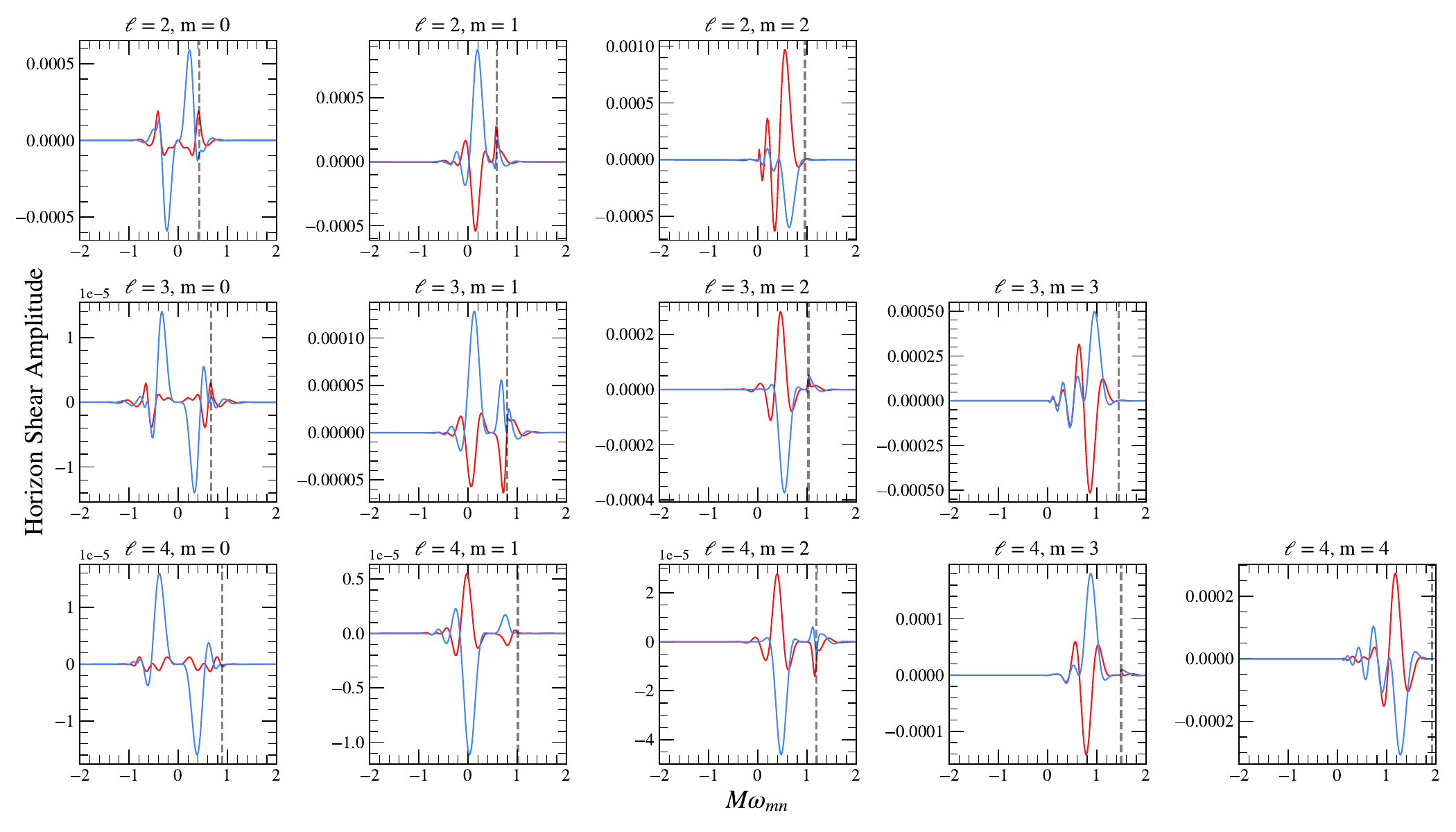}
    \caption{A panel grid of each $M^{3}\sigma_{\ell mn}$ at $a/M = 0.999$ with $p = 3.0$, $e = 0.85$, $x = 1.0$ and $(\theta,\psi) = (\pi/2, 0)$ as a function of $M\omega_{mn}$ for $\ell = 2,3,4$. Just as in Figures \ref{fig:l2m1amp} and \ref{fig:l3m2amp}, the red lines show the real part of $M^{3}\sigma_{\ell mn}$ while the blue lines show the imaginary part. The gray dashed lines show the real part of the fundamental mode for the respective $(\ell,m)$ mode at $a/M = 0.999$.}
    \label{fig:HSA}
\end{figure*}
\textcolor{black}{To be clear, by the very construction of the global phase $\Phi(v,\psi)$ in Eq.\eqref{eq:KIPhase} one sees there cannot be any QNM content in $\Phi(v,\psi)$. The only other reasonable conclusion is that it must exist in the complex amplitude}. In Figures \ref{fig:l2m1amp} and \ref{fig:l3m2amp} we show the complex amplitudes $\sigma_{21n}$ and $\sigma_{32n}$, respectively. In particular, we plot it as a function of the discrete frequency $\omega_{mn} = m\Omega_{\phi} + n\Omega_{r}$. In both cases, one can observe a sharp peak in the real amplitude and a companion slash in the imaginary amplitude superimposed with the EMRI-sourced background. These structures are visually reminiscent of a complex pole. The existence of a complex pole within the amplitudes should not be surprising. The horizon amplitude ${}_{+2}R^{\rm H}_{\ell mkn}$ is built from the near-horizon limit of the inhomogeneous Green's function. The poles of the Green's function represent the QNM solutions to the Teukolsky equation --- those in which the horizon side is purely ingoing and the infinity side is purely outgoing. The residues of these poles are the excitation factors \cite{Leaver1986}, and give an intrinsic source-independent excitation strength to each QNM \cite{Berti2006b, Hadar2011, Oshita2021, Motohashi2025, Lo2025}. In the case of an EMRI, when a tooth of the discrete frequencies $\omega_{mn}$ is sufficiently close to a pole in the Kerr Green's function, the pole becomes excited and its resulting amplitude depends on both the excitation factor and the amplitude of the EMRI-sourced structure.

In Figure \ref{fig:HSA} we plot, for $a/M = 0.999$, the horizon shear amplitudes for every $m \geq 0$ for $\ell = 2,3,4$. The $m < 0$ modes can be obtained by the symmetry
\begin{equation}\label{eq:HSAsymmetry}
    \sigma_{\ell -m-n} = (-1)^{\ell}\overline{\sigma_{\ell mn}},
\end{equation}
where the overhead bar denotes complex conjugation. From these we observe that all the $m = 0$ modes lack evidence of Kerr QNM excitation, as do all the $m = \ell$ modes, at least for $\ell = 2,3,4$. However, beyond just the $(2,1,0)$ and $(3,2,0)$ modes it appears the $(3,1,0)$, $(4,1,0)$, and $(4,2,0)$ modes exhibit pole-like structures. These three new poles are far less clear than the $(2,1,0)$ or $(3,2,0)$ poles as they are obscured by the EMRI-sourced content of $\sigma_{\ell mn}$. We therefore only include them with the interpretation of hints of further QNM content.


To analyze these poles we attempted to fit a complex pole plus polynomial background (both functions of $\omega$) to the amplitudes in Fig.~\ref{fig:HSA}, but the fitted frequencies were unstable under changes to the frequency window and often highly uncertain. The difficulty is a mismatch of scales. At a fixed $m$, the amplitudes are sampled only at $\omega_{mn}$ with spacing $\Omega_{r} \approx 0.0205$. A pole of width $|\text{Im}\omega| \approx 0.02-0.07$ has between 1 and 3 points that lie within that width, even when the real and imaginary parts are fit simultaneously. Widening the frequency window add points, but the polynomial background is only valid over ranges narrower that the oscillations of the EMRI-sourced structure whose functional form is unknown. Variation of the parameters $e$ and $p$ change the spacing of the frequency grid at the cost of decreases in the amplitude, which are in many cases already quite small.

\begin{figure*}
    \centering
    \includegraphics[width=1.0\linewidth]{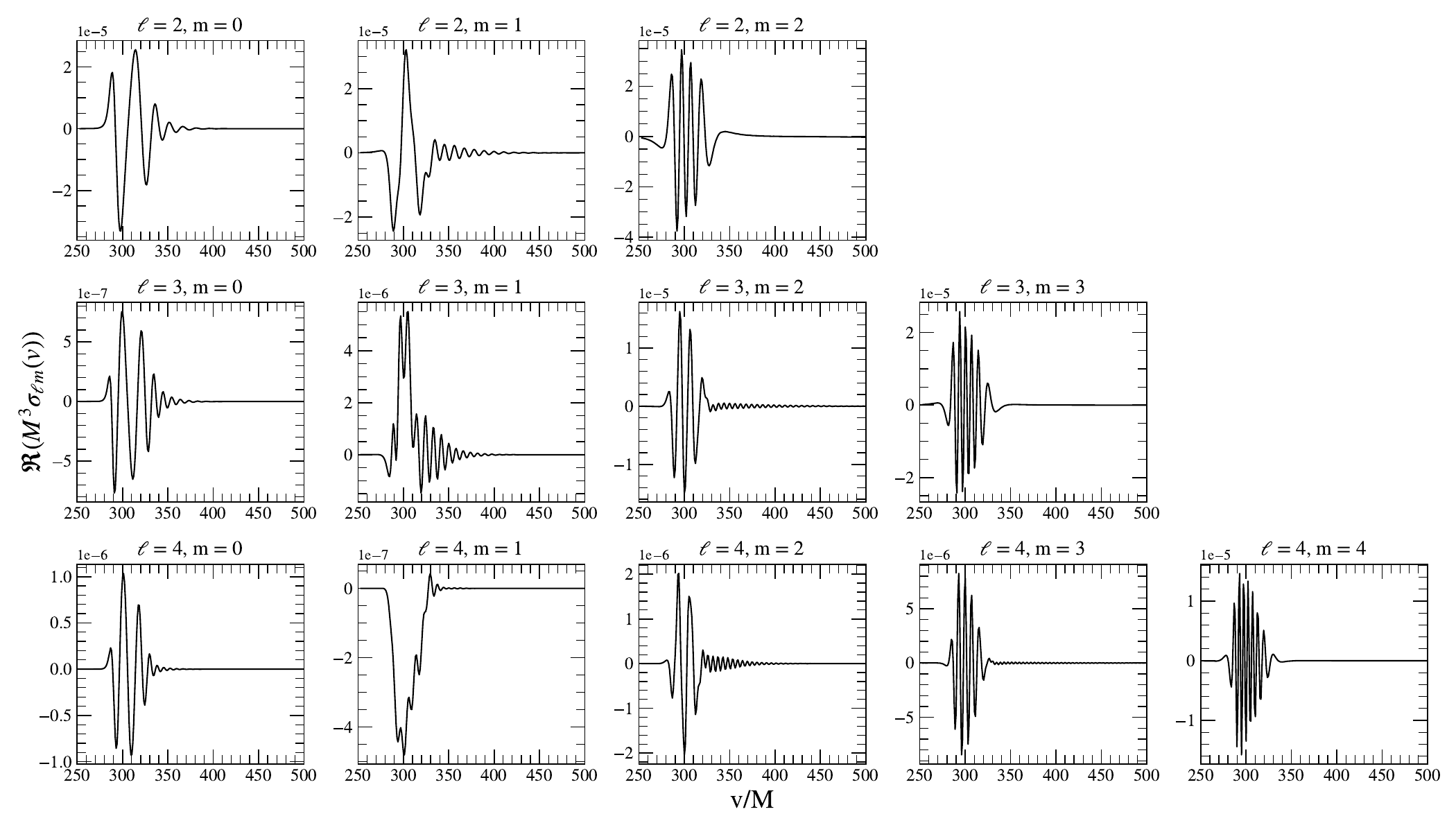}
    \caption{A panel grid of the inverse Fourier transform of each horizon shear amplitude at $a/M = 0.999$, $p = 3.0$, $e = 0.85$, $x = 1.0$ and $(\theta,\psi) = (\pi/2, 0)$ for $m \geq 0$ and $\ell = 2,3,4$.}
    \label{fig:InvFFTHorizonShear}
\end{figure*}

Instead of fitting poles in the frequency domain, we opt to work in the time domain one multipole at a time. Fixing $(\ell,m)$ in Eq.~\eqref{eq:ShearSolution} at $(\theta, \psi) = (\pi/2,0)$ gives that multipoles contribution to the shear
\begin{equation}\label{eq:FixedModeShear}
    \sigma_{\ell m}(v) = e^{-imK(a)}\sum_{n}\sigma_{\ell mn}e^{-i\omega_{mn}v},
\end{equation}
and summing $\sigma_{\ell m}$ over all $\ell$ and $m$ (with $m < 0$ given by Eq.~\eqref{eq:HSAsymmetry}) recovers the full signal of Fig.~\ref{fig:FiducialTDSignal}. Since Eq.~\eqref{eq:FixedModeShear} sums the amplitudes of Fig.~\ref{fig:HSA} over the frequency grid $\omega_{mn}$, we refer to it as an inverse Fourier transform. The motivation is that a pole at some $\omega_{\rm QNM}$ in $\sigma_{\ell mn}$ appears in $\sigma_{\ell m}(v)$ as a damped oscillation $\exp(-i\omega_{\rm QNM}v)$, while the smooth EMRI-sourced structure becomes a slowly varying background. Isolating one multipole therefore reduces the pole fit to the damped-sinusoid-plus-polynomial fit of Sec.~\ref{sec:level3.1} and is free of contamination from other multipoles. Figure \ref{fig:InvFFTHorizonShear} shows $\sigma_{\ell m}(v)$ for $a/M = 0.999$. After the burst, the $(2,1)$, $(3,1)$, $(3,2)$, $(4,1)$ and $(4,2)$ multipoles each show a high-frequency damped oscillation on top of a slower EMRI-sourced variation (for a visual aid see the $(4,1)$ multipole of Fig.~\ref{fig:InvFFTHorizonShear}, in which the aforementioned feature is particularly clear). The $m = 0,\ell$ multipoles show no such noticeable features, consistent with the apparent absence of pole-like structure in Fig.~\ref{fig:HSA}. Of the five multipoles listed above, the $(2,1)$ and $(3,2)$ multipoles have the largest raw amplitudes which is consistent with their dominance in the full time-domain signal.

\begin{table*}
\centering
    \begin{subtable}{\textwidth}
    \centering
    \renewcommand{\arraystretch}{1.3}
    \setlength{\tabcolsep}{10pt}
        \begin{tabular}{c|c|c|c|c|c|c|c|c|c|c}
            \hline
            \hline
            Spin & \multicolumn{2}{c|}{$\ell=2,m=1$} & \multicolumn{2}{c|}{$\ell=3,m=1$} & \multicolumn{2}{c|}{$\ell=3,m=2$} & \multicolumn{2}{c|}{$\ell=4,m=1$} & \multicolumn{2}{c}{$\ell=4,m=2$}\\
            \hline
            $a/M$ & $M\omega_{R}$ & $M\omega_{I}$ & $M\omega_{R}$ & $M\omega_{I}$ & $M\omega_{R}$ & $M\omega_{I}$ & $M\omega_{R}$ & $M\omega_{I}$ & $M\omega_{R}$ & $M\omega_{I}$\\
            \hline
            0.999 & 0.5808 & -0.0398 & 0.7944 & -0.0589 & 1.0257 & -0.0214 & 1.0205 & -0.0608 & 1.1909 & -0.0453 \\
            \hline
            0.9995 & 0.5815 & -0.0387 & 0.7949 & -0.0588 & 1.0266 & -0.0195 & 1.0086 & -0.0739 & 1.1916 & -0.0454 \\
            \hline
            0.9999 & 0.5813 & -0.0380 & 0.7951 & -0.0594 & 1.0217 & -0.0156 & 1.0121 & -0.0627 & 1.1917 & -0.0457 \\
            \hline
            0.99995 & 0.5810 & -0.0380 & 0.7948 & -0.0596 & 1.0265 & -0.0165 & 1.0298 & -0.0636 & 1.1909 & -0.0443 \\
            \hline
            0.99999 & 0.5800 & -0.0379 & 0.7947 & -0.0601 & 1.0247 & -0.0116 & 1.0207 & -0.0628 & 1.1910 & -0.0444 \\
            \hline
        \end{tabular}
        \caption{A table containing the fitted real ($M\omega_{R}$) and imaginary $(M\omega_{I})$ parts of the inverse Fourier transformed time-domain signals. We report findings for the $(2,1)$, $(3,1)$, $(3,2)$, $(4,1)$ and $(4,2)$ modes.}
        \label{tab:FD-A}
    \end{subtable}
    
    \vspace{1em}
    
    \begin{subtable}{\textwidth}
        \centering
        \renewcommand{\arraystretch}{1.3}
        \setlength{\tabcolsep}{10pt}
            \begin{tabular}{c|c|c|c|c|c|c|c|c|c|c}
            \hline
            \hline
            Spin & \multicolumn{2}{c|}{$\ell=2,m=1$} & \multicolumn{2}{c|}{$\ell=3,m=1$} & \multicolumn{2}{c|}{$\ell=3,m=2$} & \multicolumn{2}{c|}{$\ell=4,m=1$} & \multicolumn{2}{c}{$\ell=4,m=2$}\\
            \hline
            $a/M$ & $M\varepsilon_{R}$ & $M\varepsilon_{I}$ & $M\varepsilon_{R}$ & $M\varepsilon_{I}$ & $M\varepsilon_{R}$ & $M\varepsilon_{I}$ & $M\varepsilon_{R}$ & $M\varepsilon_{I}$ & $M\varepsilon_{R}$ & $M\varepsilon_{I}$\\
            \hline
            0.999 & 0.0008 & 0.0004 & 0.0003 & 0.0003 & 0.0004 & 0.0004 & 0.0082 & 0.0104 & 0.0072 & 0.0024 \\
            \hline
            0.9995 & 0.0008 & 0.0009 & 0.0002 & 0.0003 & 0.0008 & 0.0014 & 0.0071 & 0.0063 & 0.0003 & 0.0003 \\
            \hline
            0.9999 & 0.0009 & 0.0016 & 0.0003 & 0.0004 & 0.0051 & 0.0030 & 0.0126 & 0.0174 & 0.0005 & 0.0007 \\
            \hline
            0.99995 & 0.0011 & 0.0036 & 0.0006 & 0.0006 & 0.0034 & 0.0019 & 0.0127 & 0.0066 & 0.0010 & 0.0008 \\
            \hline
            0.99999 & 0.0014 & 0.0033 & 0.0009 & 0.0009 & 0.0056 & 0.0049 & 0.0128 & 0.0047 & 0.0008 & 0.0013 \\
            \hline
        \end{tabular}
        \caption{A table containing the uncertainties to the real ($M\varepsilon_{R}$) and imaginary $(M\varepsilon_{I})$ parts of the inverse Fourier transformed time-domain signals. We report findings for the $(2,1)$, $(3,1)$, $(3,2)$, $(4,1)$ and $(4,2)$ modes.}
        \label{tab:FD-B}
    \end{subtable}

    \vspace{1em}
    
    \begin{subtable}{\textwidth}
        \centering
        \renewcommand{\arraystretch}{1.3}
        \setlength{\tabcolsep}{10pt}
            \begin{tabular}{c|c|c|c|c|c|c|c|c|c|c}
            \hline
            \hline
            Spin & \multicolumn{2}{c|}{$\ell=2,m=1$} & \multicolumn{2}{c|}{$\ell=3,m=1$} & \multicolumn{2}{c|}{$\ell=3,m=2$} & \multicolumn{2}{c|}{$\ell=4,m=1$} & \multicolumn{2}{c}{$\ell=4,m=2$}\\
            \hline
            $a/M$ & $n_{o} = 0$ & $n_{o} = 1$ & $n_{o} = 0$ & $n_{o} = 1$ & $n_{o} = 0$ & $n_{o} = 1$ & $n_{o} = 0$ & $n_{o} = 1$ & $n_{o} = 0$ & $n_{o} = 1$\\
            \hline
            0.999 & 2.7186 & 45200 & 1.4365 & 148918 & 1.3280 & --- & 0.9613 & 195 & 0.0031 & 1678 \\
            \hline
            0.9995 & 0.3999 & 9764 & 1.3041 & 252363 & 1.6510 & --- & 1.3250 & 455 & 1.2524 & 100188 \\
            \hline
            0.9999 & 0.0626 & 9715 & 0.8387 & 73505 & 1.7718 & --- & 0.0976 & 69.8340 & 2.2241 & 15322 \\
            \hline
            0.99995 & 0.3354 & 11137 & 1.5304 & 44741 & 1.8939 & --- & 2.5384 & 519 & 2.2179 & 14869 \\
            \hline
            0.99999 & 0.9810 & 1905 & 1.7277 & 15909 & 4.2681 & --- & 0.8334 & 887 & 3.6833 & 7245 \\
            \hline
        \end{tabular}
        \caption{A table providing the Mahalanobis distances $D_{2}^{2}$ for the fitted data to their respective Kerr counterparts. We report distances to both the fundamental mode and first overtone, except for the $(3,2,1)$ mode in which we could not obtain a reliable Kerr QNM. In its place we write a dash (---). Just as in Table \ref{tab:TDFit-B}, the assignment of each mode is clear and unambiguous.}
        \label{tab:FD-C}
    \end{subtable}
\caption{A table containing the results of the time-domain fitting procedure on the inverse Fourier transformed signals for five suspected Kerr QNM excitations.}
\label{tab:FDFitResults}
\end{table*}

Tables \ref{tab:FD-A} and \ref{tab:FD-B} show the fitting results and uncertainties for the inverse Fourier transformed signal of Figure \ref{fig:InvFFTHorizonShear}, respectively. In Table \ref{tab:FD-C} we calculate the Mahalanobis distance to the respective fundamental mode and the first overtone. By isolating each multipole, we are able to obtain much more conclusive results. For the $(\ell,m) = (2,1),(3,2)$ multipoles we find that, with the exception of the $(2,1,0)$ mode at $a/M = 0.999$ and the $(3,2,0)$ mode at $a/M = 0.99999$ which are within $2\varepsilon$ of the respective Kerr QNM, each mode falls within $1\varepsilon$ of their respective fundamental Kerr QNM. We also observe close proximity to the Kerr QNM values for the other introduced multipoles, which all fall within $2\varepsilon$ of their respective Kerr QNM values.

\subsection{\label{sec:level3.3}Discussion}
\begin{figure*}
    \centering
    \includegraphics[width=1.0\linewidth]{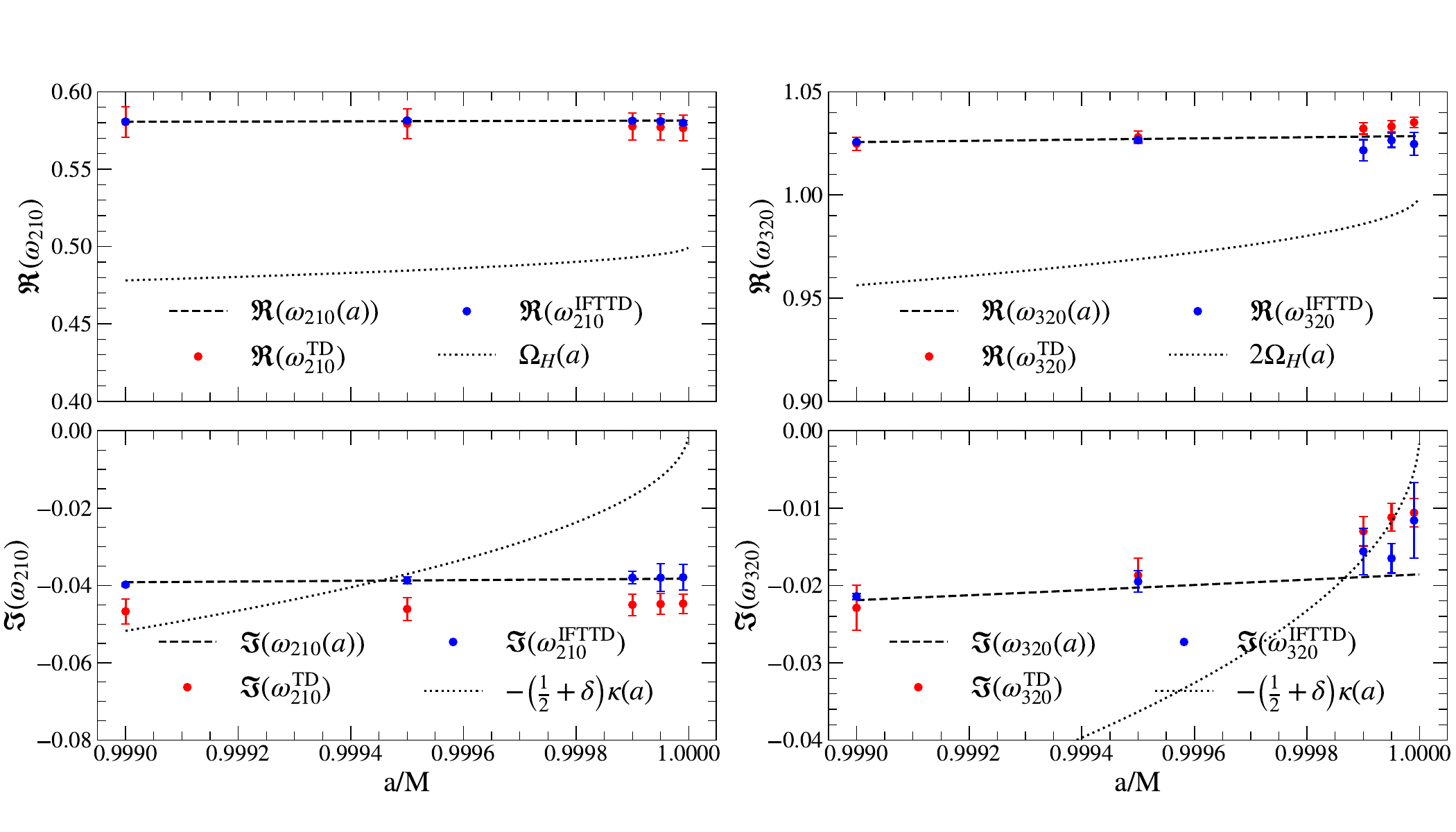}
    \caption{A plot comparing the results from Section \ref{sec:level3.1} (solid red, superscript TD) and \ref{sec:level3.2} (solid blue, superscript IFTTD) to the Kerr QNM values (dashed black) and the near-extremal ZDM values (dotted black) for the specified multipole. The alignment to the DM branch is clear, perhaps with the exception of the imaginary part of the $(3,2,0)$ mode for spins $a/M \geq 0.9999$. While it appears to trend with the ZDM curve, it cannot be a true ZDM seeing as the real part of the $(3,2,0)$ mode does not tend to $2\Omega_{H}$.}
    \label{fig:CompareFits}
\end{figure*}

At this point, we have analyzed the data in two connected ways. First, we generated the time-domain signal and fit directly to the wiggles. Then, we analyzed the horizon shear amplitudes and identified structures resembling complex poles. When inverse Fourier transforming that data, we obtained wiggles isolated by multipole and fit to that data as well. In Figure \ref{fig:CompareFits} we plot the results of these fits, and compare them to both the Kerr QNM values \textit{and} the near-extremal ZDM limits.

From Figure \ref{fig:CompareFits}, the imaginary part of the $(2,1,0)$ mode shows that fitting from the full time-domain data biases towards a shorter damping time. The likely cause is contamination from other multipole content. At early times, the signal has sufficient contributions from other multipoles which have not had enough time to decay. While in general this could either raise or lower the amplitude (depending on the relative phases), the larger $|\Im(\omega_{210}^{\rm TD})|$ as seen in Figure \ref{fig:CompareFits} suggests that the amplitude is inflated at early times. This contamination quickly subsides as their respective damping times are quite short, but this gives the appearance of a faster damping rate for the detected mode. When we isolate the $(2,1,0)$ mode, we find that the imaginary part is in close proximity with the Kerr value across all spins, and are comfortable concluding that the Kerr $(2,1,0)$ mode is excited in the horizon shear response.

The $(3,2,0)$ mode's imaginary part shows an interesting feature, a drift from the Kerr value at spins $a/M \geq 0.9999$. The results of both Section \ref{sec:level3.1} and Section \ref{sec:level3.2} show the same trend, however it is only significant in the full time-domain fit, where the deviation exceeds $3\varepsilon$. The isolated-multipole fit with its larger uncertainties stays within $2\varepsilon$. The drift is robust against the fitting procedure. Changing the fitting window, initial guess, background polynomial degree and weighting the residuals by $e^{\alpha v}$ with $0<\alpha<|\mathrm{Im}(\omega_{320})|$ all shift $\mathrm{Im}(\omega_{320})$ by no more than roughly $10^{-4}$. Including the asymptotic ZDM of Eq.~\eqref{eq:ZDMfreq} as a fixed-frequency mode gives an amplitude that is unstable under variations of the start time, indicating that it cannot be a ZDM. A four-mode fit left $\text{Im}(\omega_{320})$ essentially unchanged, but a five-mode fit moves the fitted frequency at $a/M = 0.9999$ and $a/M = 0.99999$ toward the Kerr value, potentially suggesting structure missing from our model. The origin of this structure is unclear. With only $\sim 2$ e-folds at these spins, the least-squares fit may struggle to converge, and the finite accuracy of the \textsc{qnm} package and GSN code may contribute. It does not affect our identification. The larger uncertainties of Sec.~\ref{sec:level3.2} absorb the drift, and the isolated-multipole real part agrees with the Kerr value, so we identify the $(3,2,0)$ Kerr QNM as excited.

\begin{figure}
    \centering
    \includegraphics[width=1.0\linewidth]{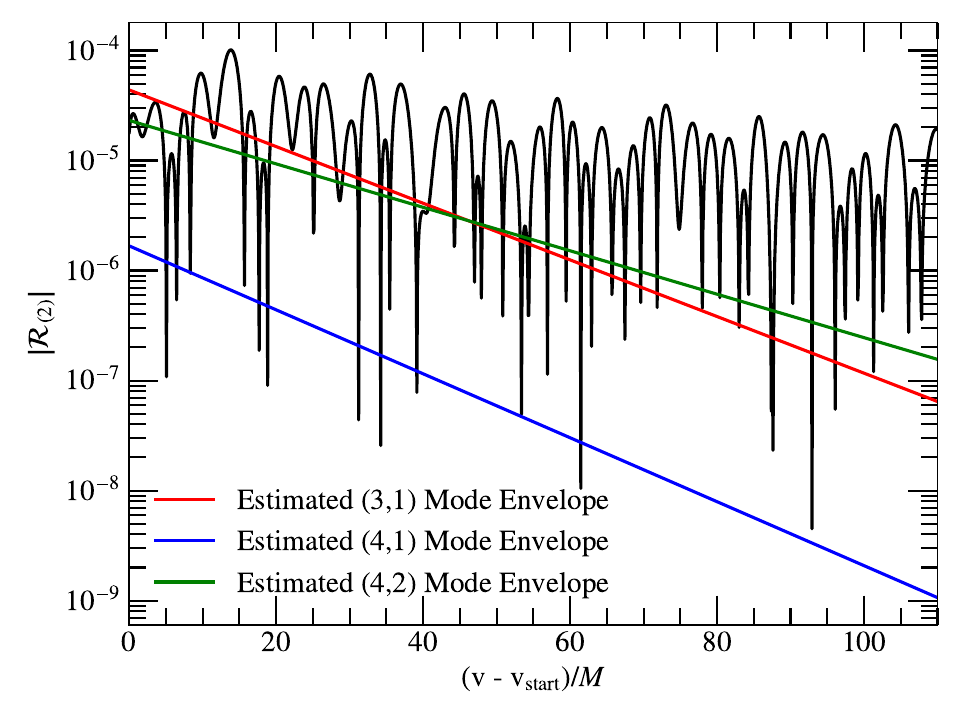}
    \caption{A plot of the estimated $(3,1,0)$, $(4,1,0)$ and $(4,2,0)$ mode envelopes --- the exponential coefficients that control the amplitude --- plotted against the two mode residuals $\mathcal{R}_{(2)}$ for $a/M = 0.999$, $p = 3.0$, $e = 0.85$, $x = 1.0$ and $(\theta,\psi) = (\pi/2, 0)$.}
    \label{fig:2ModeResiduals}
\end{figure}

Beyond identifying the $(2,1,0)$ and $(3,2,0)$ modes as excited modes, the mode isolation and identification of Section \ref{sec:level3.2} predicts, but does not confirm, the existence of the $(3,1,0)$, $(4,1,0)$ and $(4,2,0)$ modes as excitations. Ideally one could increase the number of fitted modes when fitting the full time-domain data (as in Section \ref{sec:level3.1}), but we find the added modes are sensitive to the fitting interval and therefore cannot be mode detections. In fact, one can use the fitted amplitudes obtained alongside the fitted frequencies obtained in Section \ref{sec:level3.2} to estimate the decaying envelope of these suspected excited modes. We plot these against the two mode residuals $\mathcal{R}_{(2)}$, the difference between the full data and the fit-reconstructed data, in Figure \ref{fig:2ModeResiduals}, and find that they lie at or below the residuals. While these modes may indeed be excited, they are not resolvable in our full time-domain signal without the inclusion of a large number of residual absorber modes. In Section \ref{sec:level3.2}, we only analyzed up to $\ell = 4$ while the data in Section \ref{sec:level3.1} includes up to $\ell = 9$. It is therefore possible that higher $\ell$ modes are excited, however we forego analysis of these modes because it is assumed their amplitudes would be far too small to be resolved.

This section's analysis focuses solely on the shear, but the same results apply to the tidal driving force $\psi_{0,+}^{\rm HH}$. \cite{OSullivan2014,OSullivan2016,Penna2017} mention that there is no evidence of wiggles in the tidal driving field, but the lower inset of Figure \ref{fig:FiducialTDSignal} shows the existence of these wiggles in $\psi_{0,+}^{\rm HH}$. In fact, one can argue straightforwardly that there must be wiggles in the tidal driving field using Eq.\eqref{eq:ShearTransferFunction}. Indeed, there is no QNM content in the prefactor of Eq.\eqref{eq:ShearTransferFunction}, so the existence of wiggles in the horizon shear implies the existence of wiggles in the tidal driving field and the excited QNM content is the same in each signal. Whether these wiggles are visible is a different story. Comparing the insets of Figure \ref{fig:FiducialTDSignal}, it is clear that the amplitude of the wiggles in $\psi_{0,+}^{\rm HH}$ are smaller than that of $\sigma_{+}$, which can obscure the detection of wiggles in the tidal driving field. Nevertheless, it is far more effective to use $\sigma_{+}$ to characterize the QNM content given the more prominent signal it holds.

\section{\label{sec:level4}Conclusion}
In this paper we have provided a characterization of the frequency and damping times for the wiggles observed in the horizon shear response. Using the GSN code, we generated time-domain data by discrete mode summation up to $\ell = 9$ for five different near-extremal spins. We then fit two QNMs and a quadratic background to the wiggles to obtain a measurement on the complex frequencies. For the first mode, we find the strongest agreement with the $(2,1,0)$ Kerr QNM. For all spins, this mode is within $3\varepsilon$ of the Kerr values. For the second mode, we find the $(3,2,0)$ Kerr QNM mode to be the most likely identification. While this seems to be true for $a/M = 0.999$ and $a/M = 0.9995$, at larger spins an abrupt departure from proximity to the Kerr $(3,2,0)$ mode occurs, which raised suspicion. To address this, we isolated each multipole by inverse Fourier transforming the horizon shear amplitudes and refitting the wiggles. For the $(\ell,m) = (2,1),(3,2)$ multipoles we found agreement with the Kerr QNM values to within $2\varepsilon$ for all spins. Furthermore we identified three more modes, the $(\ell, m) = (3,1), (4,1), (4,2)$ modes, which show evidence of excitation in the inverse Fourier transformed data. When comparing to the Kerr values, we also found agreement to within $2\varepsilon$, however we determined that these modes could not be resolved in the full time-domain signal.

A couple of interesting results beyond the Kerr QNM identification come from this analysis. The first is the counter-argument to the claim of no wiggles in the tidal driving field. As is shown in Figure \ref{fig:FiducialTDSignal} and discussed in Section \ref{sec:level3.3}, the wiggles are indeed present in the tidal driving field and must be consistent with the results of Section \ref{sec:level3}. Their apparent obscurity in $\psi_{0,+}^{\rm HH}$, or perhaps more appropriate their observability in $\sigma_{+}$, is due to the shear transfer function in Eq.\eqref{eq:ShearTransferFunction}. While in theory it is possible to analyze $\Psi_{0,+}^{\rm HH}$ for wiggles, we find it much more appropriate to use $\sigma_{+}$ given that its signal, relative to the background, is far stronger than that of $\Psi_{0,+}^{\rm HH}$.

Another interesting observation is the lack of QNM DM excitation in the $m = 0$ mode. For the $m = \ell$ mode this is expected. With the definition $\gamma = m/\left(\ell + \frac{1}{2}\right)$, \cite{Yang2013a,Yang2013b} find that for $\gamma \lesssim 0.74$ both the ZDM and DM branches exist. So, while DMs are not present in the $m = \ell$ multipoles, there is nothing forbidding their existence in the $m = 0$ mode, and yet there are no hints of DM excitation. We currently have no explanation for this and leave it to be investigated in future work.
\acknowledgements
The author is grateful to Richard Price and Zhen Pan for their insightful feedback and revisions of this manuscript. This work makes use of the Black Hole Perturbation Toolkit \cite{BHPToolkit}.
\bibliography{bibliography}
\end{document}